\documentclass[AMA,STIX1COL]{WileyNJD-v2}

\articletype{Article Type}%

\received{2019}
\revised{}
\accepted{}

\usepackage{listings, xcolor}

\definecolor{verylightgray}{rgb}{.97,.97,.97}

\lstdefinelanguage{Solidity}{
	keywords=[1]{anonymous, assembly, assert, balance, break, call, callcode, case, catch, class, constant, continue, contract, debugger, default, delegatecall, delete, do, else, emit, event, export, external, false, finally, for, function, gas, if, implements, import, in, indexed, instanceof, interface, internal, is, length, library, log0, log1, log2, log3, log4, memory, modifier, new, payable, pragma, private, protected, public, pure, push, require, return, returns, revert, selfdestruct, send, storage, struct, suicide, super, switch, then, this, throw, transfer, true, try, typeof, using, value, view, while, with, addmod, ecrecover, keccak256, mulmod, ripemd160, sha256, sha3}, % generic keywords including crypto operations
	keywordstyle=[1]\color{blue}\bfseries,
	keywords=[2]{address, bool, byte, bytes, bytes1, bytes2, bytes3, bytes4, bytes5, bytes6, bytes7, bytes8, bytes9, bytes10, bytes11, bytes12, bytes13, bytes14, bytes15, bytes16, bytes17, bytes18, bytes19, bytes20, bytes21, bytes22, bytes23, bytes24, bytes25, bytes26, bytes27, bytes28, bytes29, bytes30, bytes31, bytes32, enum, int, int8, int16, int24, int32, int40, int48, int56, int64, int72, int80, int88, int96, int104, int112, int120, int128, int136, int144, int152, int160, int168, int176, int184, int192, int200, int208, int216, int224, int232, int240, int248, int256, mapping, string, uint, uint8, uint16, uint24, uint32, uint40, uint48, uint56, uint64, uint72, uint80, uint88, uint96, uint104, uint112, uint120, uint128, uint136, uint144, uint152, uint160, uint168, uint176, uint184, uint192, uint200, uint208, uint216, uint224, uint232, uint240, uint248, uint256, var, void, ether, finney, szabo, wei, days, hours, minutes, seconds, weeks, years},	% types; money and time units
	keywordstyle=[2]\color{teal}\bfseries,
	keywords=[3]{block, blockhash, coinbase, difficulty, gaslimit, number, timestamp, msg, data, gas, sender, sig, value, now, tx, gasprice, origin},	% environment variables
	keywordstyle=[3]\color{violet}\bfseries,
	identifierstyle=\color{black},
	sensitive=false,
	comment=[l]{//},
	morecomment=[s]{/*}{*/},
	commentstyle=\color{gray}\ttfamily,
	stringstyle=\color{red}\ttfamily,
	morestring=[b]',
	morestring=[b]"
}

\begin{document}

\title{Using blockchain in the follow-up of emergency situations related to events}

\author[1]{Alexandra Rivero-Garc\'ia*}
\author[1]{Iv\'an Santos-Gonz\'alez}
\author[1]{Candelaria Hern\'andez-Goya}
\author[1]{Pino Caballero-Gil}

\authormark{Alexandra Rivero-Grac\'ia \textsc{et al}}

\address[1]{\orgdiv{Departament of Computer Science and Systems}, \orgname{University of La Laguna}, \orgaddress{\state{Tenerife}, \country{Spain}}}

\corres{*Alexandra Rivero Garcia. \email{ariverog@ull.edu.com}}

%\presentaddress{This is sample for present address text this is sample for present address text}

\abstract[Summary]{This paper describes a decentralized low-cost system designed to reinforce personal security in big events in case of emergency.
The proposal consists of using  smart contracts supported by blockchain in the management of events. An alternative communication channel that does not require any cloud service is also provided with the aim of improving the coordination of emergency services.
Peers may use this emergency support tool to interact with each other through a chat when additional support is required. 
Since information security is mandatory in this scenario, Identity-Based Signcryption schemes are here used in order to guarantee communication confidentiality, authenticity and integrity.
Depending on the communication mode (peer-to-peer or broadcast), different signcryption methods are used. A first implementation of the proposal has produced promising results.}


%\jnlcitation{\cname{%
%\author{Williams K.}, 
%\author{B. Hoskins}, 
%\author{R. Lee}, 
%\author{G. Masato}, and 
%\author{T. Woollings}} (\cyear{2016}), 
%\ctitle{A regime analysis of Atlantic winter jet variability applied to evaluate HadGEM3-GC2}, %\cjournal{Q.J.R. Meteorol. Soc.}, \cvol{2017;00:1--6}.}

\maketitle

%\footnotetext{\textbf{Abbreviations:} XXXXXXX}

\section{Introduction} \label{ar.sec.estadodelarte}

%% MOD 1 - 2 --> 
The number of massive events in the cities is constantly increasing due to flood risk, protest march, a concert, a fire, etc. and all of them has to be controlled by the rescue staff (police, firefighters, medical staff, etc.). The fast evolution of the communication technologies has generated multiple new approaches in different scenarios where the use of smartphones to backing the different daily tasks, due to their small size and high performance, is more important every day.
One of the most important things of this small tools is that they are equipped with very powerful communication technologies that can help in different scenarios, including the aforementioned. 
%% MOD <-- 

The emergence of  blockchain technology has driven the introduction of decentralized data structures in multiple scenarios. Its main point of interest is the possibility of storing huge amounts of data in network nodes, enabling them to verify and approve any transaction. Integrity protection of the information stored in this data structure is also guaranteed.

This paper presents a decentralized low-cost model based on blockchain and on the establishment of an alternative communication channel to improve the intervention of emergency services without requiring any cloud service. %that try to improve, on the one hand, the generation of events by the different kind of people who participate in the rescue staff based on blockchain and smart contracts; and on the other hand, the communication of the rescue staff in scenarios where network congestion is produced by massive access of users involved in emergency situations.
In particular, the use of a permissioned blockchain is a fundamental issue in this proposal because write permission is granted only to qualified members while the information generated for an event may be accessible and verifiable by all personnel involved in it, at any moment. 
In particular, only authorized staff can read the blocks, execute a smart contract and verify new blocks. 
%% MOD 2 --> 
This approach facilitates coordination among emergency services, because thanks to this if the event is generated by someone that below to an specific organization, the rest of the organizations can verify in every moment the generated event without the necessity of a specific global system. 
Moreover, everyone can access to the information of the event and manage the information for the coordination in a private way supposing it an important advantage in relation to the process used nowadays.  
%% MOD <-- 

The idea behind the proposal  is to associate incidents with blocks in a smart contract. Once any member of emergency organizations detects an incident, a new block is generated to be include in the blockchain after been validated by nearby staff. Then, alerts are issued to the rest of the assigned emergency staff. As a result of this, the emergency staff has access to the event information and may act depending on it.
The smart contract has to be created by some authorized member of an emergency  body. Initially, formatted information regarding the event and the assigned emergency staff is sent to the smart contract.
Part of this information is related to the security of the communication among workers, based on the event information.
Once the event is in the blockchain, the first step is the assignment of different emergency service resources to specific areas to help to preserve civil security.

\begin{figure}[!ht]
\centering
\includegraphics[width=0.8\linewidth]{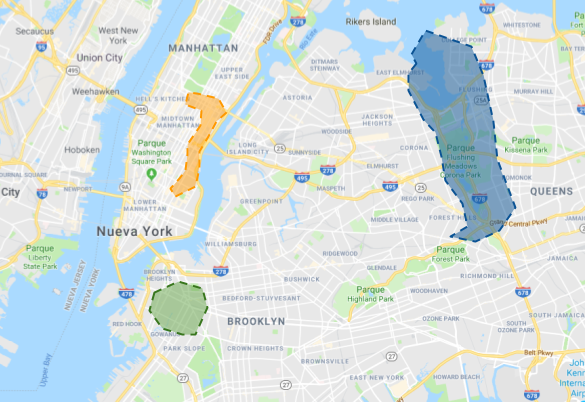}
\caption{Geolocation of events.} \label{ar.fig.events}
\end{figure}

%% MOD --> 
We can see an example in Figure \ref{ar.fig.events} where there are three simultaneous events: a cultural event with large flow of people in green, a protest march in orange and an area with high risk of flood in blue. 
All this events are verified and different types of emergency service workers must be assigned to the different zones.

% MOD <-- 

When an event is generated, the assigned emergency staff may access the block and some pre-shared information related to the event.
Based on this information, workers can participate via mobile phone in a generic event chat associated to the specific event to share information related to coordination.

Currently, communications among emergency services are carried out mostly by radio frequency.
Here, the use of two different wireless technologies through smartphones: Bluetooth Low Energy (BLE) \cite{gupta2016inside} and Wi-Fi Direct \cite{shen2016secure} is proposed.
In cases of network congestion, the system will declare ``emergency mode'' and the communications will proceed directly through these technologies in  registered smartphones. Two communication modes are supported: Peer-TO-Peer (P2P) mode, where the system establishes a direct channel through Wi-Fi Direct between two registered participants; and broadcast mode, where the system shares a message through BLE with all the recipients simultaneously.

Since this work deals with critical situations, communication security is essential.
That is why an identity-based encryption scheme is here proposed \cite{ar.Boneh2001}. Specifically, an identity-based signcryption scheme based on the geolocation and the public identification of emergency service workers is used. All  shared messages are signed and encrypted with this scheme.
The used signcryption scheme  is a combination of an ID-Based Signcryption Scheme\cite{malone2002identity}
and an ID-Based Signcryption Scheme for Multiple Receivers \cite{Sharmila2008Efficient}.     

%% MOD --> 1
In the proposal, communication is done through two different technologies using smartphones: Bluetooth Low Energy (BLE) \cite{gupta2016inside} and via Wi-Fi Direct \cite{shen2016secure}.
The features described below will be taken into account to choose the alternative. 
When possible the channel created by Wi-Fi Direct, due to its higher rate of speed and its greater range, will be used. Bluetooth Low Energy has a transmission rate of 25 Mbps and Wi-Fi Direct has a transmission rate of 250 Mbps. The maximum range of Bluetooth Low Energy Communication is 60 meters, while Wi-Fi Direct has a range of 200 meters. In the same range of Wi-Fi, Wi-Fi Aware improves the performance of Wi-Fi Direct. Wi-Fi Aware \cite{das2019radio} is only available for the latest version of Android \cite{android2011android} and as a preview mode. 
%% MOD <-- 

This paper is  structured as follows.
Section \ref{ar.sec.related} includes a review of publications related to the proposed system.
In Section \ref{ar.sec.preliminaries} some preliminaries are explained while the proposed system is introduced in Section \ref{ar.sec.globalview}.
The event generation using blockchain-based smart contracts is defined in Section \ref{ar.sec.ibs}, jointly with the details of the used communication scheme and its formal description. 
Section \ref{ar.sec.anali} deals with the description of the elements included in the implementation. 
Finally, Section 7 ends the paper, summarizing the main contributions of the proposal.

%%%%%%%%%%%%%%%%%%%%%%%%%%%%%%%%%%%%%%%%%%%%
\section{Related works}\label{ar.sec.related}
%%%%%%%%%%%%%%%%%%%%%%%%%%%%%%%%%%%%%%%%%%%%

This paper describes an application of smart contracts in the secure management of emergencies in large events. Today, most communications deployed in emergency situations are based on RF technology, which can be considered a poor solution because it only allows audio to be shared on a specific frequency, and it is not possible to group and share media data effectively.

%% MOD --> 7
On the one hand, some solutions based on the modeling and evaluation of emergency management support system has been created. In \cite{dOroCGIMP19} a use of case based on a complex edge computing is proposed. 
%% <-- MOD
On the other hand, multiple solutions based on Wi-Fi Direct have been proposed, such as \cite{motta2010wireless} where the potential of Wi-Fi Direct in the implementation of mobile P2P systems is evaluated.

%% MOD --> 
That work includes some examples of the use of Wi-Fi Direct to share text messages, to disseminate information.  A middle-ware for P2P networking is used to distribute hash tables to search for peers.
The work  \cite{conti2013experimenting} proposes generating opportunistic networks over Wi-Fi Direct by studying the latency at the link layer. 
That is an extension of \cite{camps2013device}, where multiple groups are generated and experimental measurements are  presented to confirm the suitability of Wi-Fi Direct for P2P systems. 
%% MOD <-- 

The use of Wi-Fi Direct for  alternative communication  in emergency situations was proposed in \cite{santos2014alternative}, but not for communication among emergency services. The main goal of that application is to share the geolocation of  victims when they are isolated. 
A first implementation may be found at \cite{rivero2018secure}, where a partial solution for communication in emergencies according to a centralized model was proposed. 

The approach here described differs from others in that it takes into account the distribution, assignment and location of human resources in big events, and information security is addressed as a global requirement. Blockchain is included as an specific tool to address security aspects.
 
Regarding blockchain background, the first application of the concept was in the finance scenario \cite{nakamoto2008bitcoin}. A decentralized model to share information, including the concept of transaction is also described there. 

Two other more recent contributions to the health and sanitary setting using blockchain are \cite{ekblaw2016case} and \cite{dubovitskaya2017secure}.
The first one describes a prototype to allow patients to have access to all their medical data through the  integration of data and medical providers according to different roles. The second paper aims to improve the process of data sharing between researchers and health care providers.

The work \cite{Radanovic2018} provides a complete analysis of the use of blockchain in medicine.
Although it points out scalability and data access control as general obstacles, it identifies several promising groups of applications in that scenario. For instance, the  management of electronic health records is highlighted as one of these areas of interest.

Some initiatives, such as Hyperledger, have been used to develop different solutions based on blockchain for healthcare and Internet of Things (IoT).
%% MOD --> 7
A very detailed analysis of the use of IoT and blockchain has been created on the work \cite{MinoliO18}. There, the authors analyze all the blockchain mechanisms for IoT security.  
%% <-- MOD
The work \cite{cha2018blockchain} presents a solution based on the monitorization of IoT sensors in a connected gateway for BLE. Patients may control which users have access to their data through a mobile application.

In the work \cite{guo2018secure}, a secure attribute-based signature scheme is proposed to store medical health records in a blockchain and to manage the access to this information based on the attributes of each user. There, multiple authorities are allowed to access to the blockchain, limiting the accessible data set. 
In a similar way, the authors of \cite{yue2016healthcare} propose the use of blockchain to preserve privacy of medical data.

Another solution based on the use of smart contracts in healthcare is \cite{griggs2018healthcare}, where patients are monitored to detect events related to medical conditions. This system allows sending notifications when conditions change and a record of activities is also provided.

%%%%

%%%%%%%%%%%%%%%%%%%%%%%%%%%%%%%%%%%%%%%%%%%%
\section{Preliminaries} \label{ar.sec.preliminaries}
%%%%%%%%%%%%%%%%%%%%%%%%%%%%%%%%%%%%%%%%%%%%

\subsection{Blockchain}

    Roughly speaking, blockchain is a decentralized database that stores a registry of assets and transactions across a computer network. Specifically, it can be seen as a P2P network that is secured through  strong cryptography.  Each item holds a timestamp and a link to a previous document. In this manner, once this item is sealed, it is theoretically impossible to modify it. Hence, the information inserted in the blockchain is persistent once it is inserted in the system. 
    
    A transaction takes place when the timestamp is obtained. This procedure provides  the blockchain with a time registration mechanism, enabling the possibility of knowing the timeline of information generation.
    
\begin{figure}[!ht]
    \centering
    \includegraphics[width=0.9\linewidth]{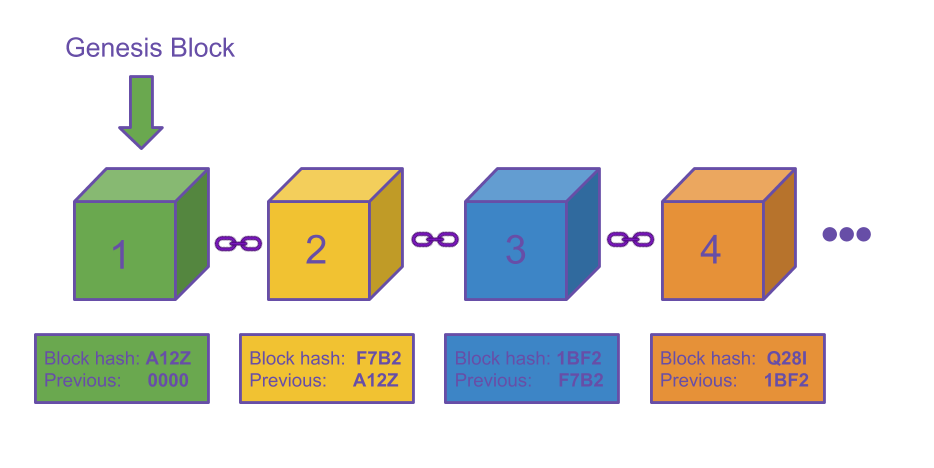}
    \caption{Blockchain example.} \label{ar.fig.blockchain}
    \end{figure}

    Blocks represent confirmed transactions. Each block  contains a code linking it to the previous block, some information related to the transaction (involved people, amount of cryptocurrencies, etc.), and another code linking it to the next block. Both codes are computed with a hash function used to generate a chain.
    A basic example is shown in Figure~\ref{ar.fig.blockchain}, where relations between blocks 
    according to their hash codes is illustrated.
    
    There is a main block that is the first one named genesis block.
    This block is identified because the hash value of its previous block is '0000'. 
    In a blockchain, a token is a representation of a physical or digital asset built on some native currency.
    Blockchain tokens include some specific properties like a name, a symbol, the initial number of minted units, the maximum number of units, the severability (because a token can be divisible into smaller units or be indivisible) and a link either to an underlying physical or virtual value or asset, or to give it power to perform some action.

% MOD --> 2 , 3
   % The elements of the blockchain are replicable on different computers, since what underlies  is a distributed database. This justifies that it is impossible to falsify its elements, since all the data are generally found in several servers, and its synchronization occurs almost simultaneously. In addition, if a falsification is achieved with one of the registers, it should be easy to detect it through the codes that link the blocks and prevent it from spreading.
    
    %Before including a transaction into a block it is always verified. This verification is carried out by miners, which are computers dedicated not only to keep a copy of the data but also to validate nodes' agreement. 
% MOD <--

\subsection{Smart contracts }

   A contract is an agreement between two or more parties, including a set of requirements and execution conditions accepted by the endorsers. Up to now, contracts have been  written documents subject to the laws.
   Smart contracts extend this concept to computer programs that execute agreements established between two or more parties when a pre-programmed condition happens.
 
 % MOD --> 3
    %That is, they are contracts that are executed and enforced  automatically and autonomously without the intervention of third parties.
    %A smart contract can be created and executed by individuals and/or legal entities, but also by machines or other programs that work autonomously.
% MOD <-- 

    Due to its nature, a smart contract is valid without the need of authorities because it is a code. Blockchain technology makes possible to share this code with all network nodes, guaranteeing that it cannot be modified. Hence, the main features of smart contracts are decentralization, persistence and transparency.
 
 % MOD --> 3
    %The so-called oracles are  elements that allow the smart contract to interact with the real world. They are autonomous tools that allow updating the internal states of a smart contract through external information, usually obtained with APIs.
    %In the proposal here described, oracles are smart devices used by the emergency services.
% MOD <-- 
    
    Many of the latest implementation proposals for smart contracts are based on Ethereum. This option operates on a distributed computing platform based on a public blockchain using a crypto-currency called ether.
    Solidity is the language used in Ethereum. It is a statically typed object-oriented language similar to Java in its syntax.

\subsection{Identity-based signcryption} 

    Identity-Based Encryption (IBE) avoids typical problems related to certificates in public key cryptography because public keys are information  used to identify  users of the system (email address, social security number, personal identifier, etc).
    Thus, the use of an IBE scheme allows the user to encrypt data with the public identifier associated to the the receiver, and to decrypt the message, the recipient will use private information that only he/she knows.
        %\cite{ar.Boneh2001}.
        %\cite{joye2009identity}
        %There are multiple solutions based on that like \cite{li2015identity}, \cite{hohenberger2013full} and \cite{wang2015identity}.
        
    Based on this idea, some variants can be found in the scientific literature. For example, Identity-Based SignCryption (IBSC) schemes where a composition of an encryption scheme with a signature scheme is defined.
    This combination allows the system to guarantee integrity, confidentiality, authenticity and non-repudiation efficiently and in a single step.
    
    The best known IBSC schemes are based on bilinear pairings on elliptic curves.
    
        %\cite{boyen2010identity}.
        %There are several proposals like in \cite{li2013fully}, \cite{li2013secure} and \cite{libert2003new}.

\subsection{Bilinear pairing} 

A bilinear pairing can be defined as follows. Let $( \mathbb{G}, +)$ and $(\mathbb{G_T}, \cdot)$  be two cycling groups of the same prime order $q$. 
	Let $P$ be a generator of $ \mathbb{G}$ and $\hat{e} : \mathbb{G} \times \mathbb{G} \rightarrow \mathbb{G_T}$  be a bilinear map pairing that satisfies the following conditions:

    \begin{itemize}

        \item \textit{Bilinear}, $ \forall P, Q \in \mathbb{G}$ and $\forall a, b \in \mathbb{Z}$, $\hat{e}(aP, bQ) = \hat{e}(P, Q)^{ab} $.

    	\item \textit{Non-degenerate}, $ \exists P_1, P_2 \in \mathbb{G} $ such that $\hat{e}(P_1,P_2) \neq 1$. 
    	This means that if $P$ is a generator of $\mathbb{G}$, then $\hat{e}(P,P)$ is a generator of $\mathbb{G_T}$. 

    	\item \textit{Computability}, there exists an algorithm to compute $\hat{e}(P,Q), \forall P,Q \in \mathbb{G}$. % \cite{groth2008efficient}.
        
    \end{itemize}

\subsection{Elliptic Curve Discrete Logarithm Problem} 

    Let's consider the cyclic group \{$ 0 , P, 2P, 3P, ... $\} for any point $P$ on an elliptic curve where the operation $kP$ is called scalar multiplication, being $k$ an integer.  
    The Elliptic Curve Discrete Logarithm Problem (ECDLP) consists in finding $k$, given the points $kP$ and $P$.
    Solving the ECDLP for appropriate parameters is computationally infeasible, and is the basis of the proposed scheme.

%%%%%%%%%%%%%%%%%%%%%%%%%%%%%%%%%%%%%%%%%%%%
\section{Global System View} \label{ar.sec.globalview}
%%%%%%%%%%%%%%%%%%%%%%%%%%%%%%%%%%%%%%%%%%%%

\begin{figure}[!ht]
	\centering
	\includegraphics[width=0.9\linewidth]{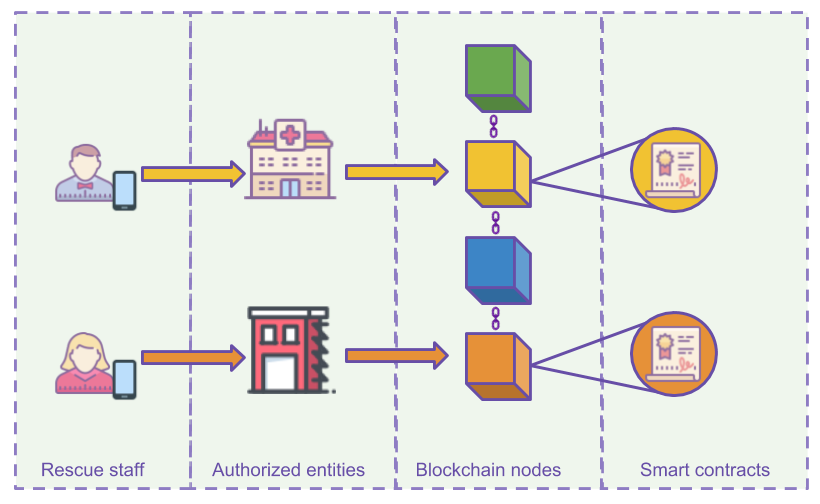}
	\caption{Block generation process.} \label{ar.fig.smartcontracts}
\end{figure}

The main idea behind the proposal is to put forward  a permissioned blockchain to monitor risk level in big events and to improve the deployment of material and human resources if an emergency occurs. 
Permissioned blockchains are those in which transaction processing is carried out by a list of known and authorized participants (Figure \ref{ar.fig.smartcontracts}).

Participant entities may be hospitals, emergency services, police stations, health centres, fire stations and forest guards and other governmental authorities, such us municipalities, local, state or national authorities.

The process presented here starts when a member of an emergency body detects an incident and generates a notification by using his/her smartphone to be send to his/her organization. 
This notification contains the geolocation, the kind of incident, and the estimated level of risk.  
The proposed signcryption scheme is applied on this notification to guarantee its confidentiality, integrity and authenticity.  

\begin{figure}[!ht]
	\centering
	\includegraphics[width=1\linewidth]{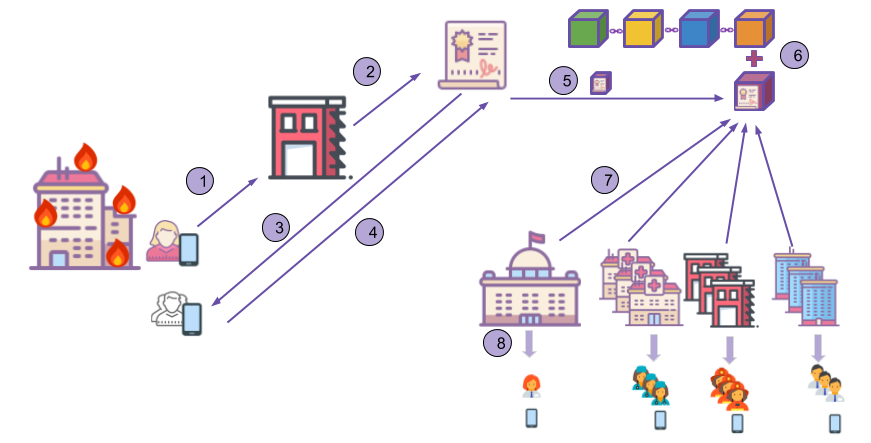}
	\caption{Flow of the event generation.} \label{ar.fig.flowevent}
\end{figure}

Once the corresponding entity verifies the identity of the person who sent the notification, it generates the smart contract and requests ratification from other members belonging to the organization located near the incident.
If any ratification is received, a new block is generated and inserted into the blockchain with the corresponding output of the smart contract.
Consequently, all the participating entities may access to the incident information through the blockchain (see Figure \ref{ar.fig.flowevent}).  

After adding the incident to the blockchain, resource allocation to limit  possible risks is carried out.
The resource inventory assigned to the incident is made publicly available by including it in the smart contract.

The system uses a permissioned blockchain to distribute keys for secret communication during the event. The personnel may access the blocks, and a chat service where they may participate because  they have  some pre-shared information, is enabled for the event.
Thus, the participants do not need any native token because the assets are the specific resources (staff, machinery, etc.) owned by the entities. %% MOD --> 6
To make it possible, the created network  has to have at least the 50\% of the users of the blockchain to have  consensus and write new blocks. This is possible because this blockchain is used only in the emergency and most of the users are there.
 %% MOD <-- 6

In order to prevent network congestion, an alternative communication system for emergency service staff and supported by mobile phones is provided. It supports two different communication modes: P2P and broadcast. 
%More details are given in section \label{ar.sec.ibs}.

\begin{figure}[!ht]
    \centering
    \includegraphics[width=0.65\linewidth]{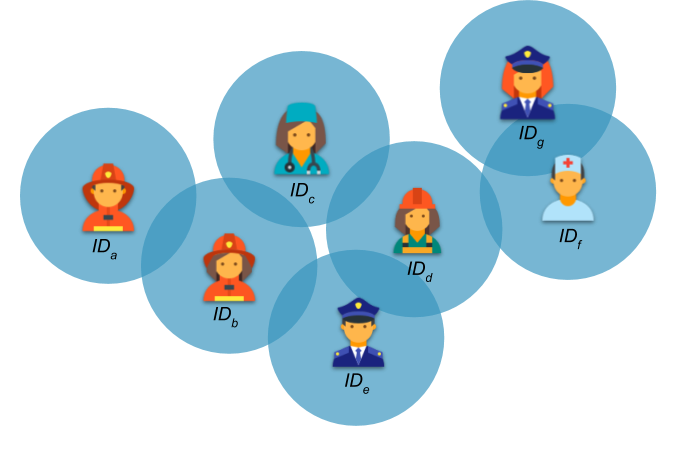}
    \caption{Sharing identifiers through beacon mode.} \label{ar.fig.beacon}
\end{figure}

When the emergency mode is activated, it is necessary to share some users' public information regarding user's ID. This information is shared here  through BLE using beacon mode (see Figure \ref{ar.fig.beacon}).
Every participant has a list of identifiers (IDs) corresponding to nearby people. This list must be made publicly available in order to verify who the legitimate participants are. This is achieved by including such a list in the smart contract.

%In the same range of Wi-Fi, Wi-Fi Aware improves the performance of Wi-Fi Direct. 
%Wi-Fi Aware \cite{schilit2003challenge} is only available for the latest version of Android \cite{android2011android} and as a preview mode. 

Data shared through this new communication channel must be protected. An ID-Based Signcryption scheme is implemented to complete this task in both communication modes. Participating in secure communications is possible only when possessing an identifier included in the smart contract list. 

There is a central application (mobile application) to handle the event and distributing information among the participant staff. Its management is collaborative and all the entities involved in the event may participate in it. The initialization steps are as follows. Firstly, the authorized entity generates the event and assigns different types of resources to the event. Specific information that allows staff participation in the chat system is also provided.

A unique identifier randomly generated is assigned to each event together with its geolocation. 
In order to prevent the generation of false multiple events, it is considered that a range of some miles refer to the same event.   
When a member of the emergency staff is assigned to an event, the system generates specific credentials and keys to share data.
Users may get from the mobile application their own location, theirs peers' locations and the  scope area of the event.

\begin{figure}[!ht]
    \centering
    \includegraphics[width=0.9\linewidth]{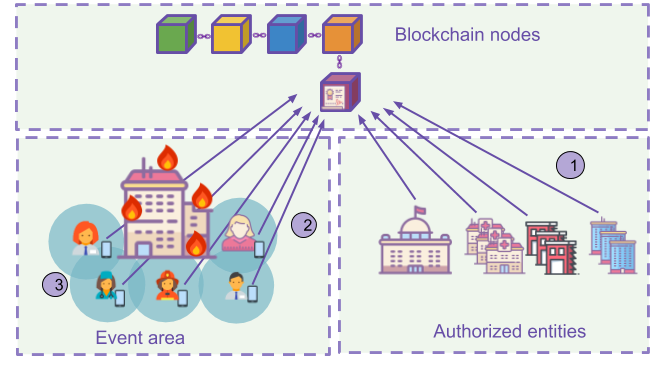}
    \caption{Information access.} \label{ar.fig.flowevent2}
\end{figure}

The system is designed to allow that emergency staff with heterogeneous communication capabilities may interact and  have access to shared information. 
The event information can be updated and sent to the blockchain as many times as needed by generating a new block containing the reference to the event identifier (eventID).
Figure \ref{ar.fig.flowevent2} describes how the information may be accessed.   
 
Among  the relevant fields included in the smart contacts we have the identifier of the person who generates the event (generator), the entity to which this person belongs, and the rules (privacyPolicy) to access the smart contract. Apart from that, the system defines some values like the identifier of the event (eventID) and its geolocation (location).

Events are classified by using a code in the contract denoted as kind (fire, climate phenomenon, seismic phenomenon, volcanic phenomenon, flooding, pollution, etc.) in order to estimate what resources it requires. State represents the status of the event: created (refers to the state in which a person sends the creation of the event before receiving confirmation from the rest of the participants), verified (is used when a real event exists and some staff is working in it) and inactive (is the state used when an event is finished). 
%% MOD --> 1
\textit{State} represents the status of the event: 
\begin{itemize}
	\item Created, refers to the state in which a person sends the creation of the event but there is still no confirmation from the rest of the participants.
	\item Verified,  is used when a real event exists where some staff is working.
	\item Inactive,  is the state used when a event is finished. 
\end{itemize}
%% MOD <-- 

There are two fields that refer to the staff assigned to the event: participants stores the identifiers of the people who participate in the event, and  numParticipants stores the number of participants.
Note that each user has an identification (IDentity) associated to his/her own address (user) and also an entity to which he/she belongs.

Related to the event generation, the contract also includes when the event was generated (EventGeneration()). If  people nearby confirm the event, the contract includes (EventConfirmed()); otherwise, it includes  (EventAborted()).

Several functions related to the internal operation of the smart contract and the evolution of the event exist.  One of the most important fields for the communication system is getIDs(), which is used to disseminate staff public identification. 
Besides, getSharedData() allows users to generate the communication with some specific pre-shared security information. 
An example of the smart contract used in the proposed system is shown in Algorithm \ref{ar.pseudo.sc}.

\begin{lstlisting}[language=Solidity, caption={Pseudo-code of the smart contract to generate an event}, label={ar.pseudo.sc}, float={!ht}]
pragma solidity 0.4.16;

contract Event {
	address entity;
	address generator;
	address public privacyPolicy;
	string public eventID;
	string public location;
	string kind; 
	uint riskLevel;
	enum State { Created, Verified, Inactive }
	State public state;
	Worker[] public participants;
	uint public numParticipants;
	struct Worker{
		address entity;
		address user;
		string IDentity;
		string enventID;
	}	
	...
	modifier onlyEntity(){...}
	
	event EventGeneration(eventID) {...}
	event EventConfirmed();
	event EventAborted();
  
	function Event onlyEntity (string _eventID, address _privacyPolicy, string _location, string _kind, uint _riskeLevel) {...}
	function UpdateParticipants(string _eventID, address _privacyPolicy, Workers[] _participants, uint numParticipants, Worker _participant, address _entity) {...}
	function UpdateState(string _eventID, uint _riskeLevel, State _state) {...}
  function UpdateAccess (string _eventID, address _privacyPolicy) {...}
	function kill onlyEntity (string _eventID) {...}
	function getIDs (string _eventID, address _privacyPolicy, Worker _participant) {...}
	function getSharedData (string _eventID, address _privacyPolicy, Worker _participant) {...}
	...
}

\end{lstlisting}

%%%%%%%%%%%%%%%%%%%%%%%%%%%%%%%%%%%%%%%%%%%%
\section{Emergency communication scheme}\label{ar.sec.ibs}
%%%%%%%%%%%%%%%%%%%%%%%%%%%%%%%%%%%%%%%%%%%%

As it has been mentioned before, an IBSC scheme is used with the aim of guaranteeing communication security.
%This approach offers the advantage of simplifying management by not having to define a public key infrastructure.
The rationale behind this selection was its low computational complexity and its high efficiency in terms of memory, and usability. 

This communication system allows  sharing text, images and audio. Emergency service staff is provided with two variants of IBSC  to share information. When the information exchange involves P2P mode, an ID-Based Signcryption is used. Otherwise, in broadcast mode, an ID-Based Multi-Receiver Signcryption Scheme is implemented.

The features described below were taken into account when choosing the communication technology to be used. 
The channel is supported by Wi-Fi Direct whenever possible, due to its higher speed rate and greater range. A second option is  BLE technology, where the  transmission rate is of 25 Mbps (while Wi-Fi Direct has a transmission rate of 250 Mbps).
Another important difference between these technologies that justifies the previous preference refers to the maximum range of BLE, which is 60 meters, while Wi-Fi Direct has a range of 200 meters.

% MOD --> 5 
In order to define this communication system it is required to share some users' public information when emergency mode is activated. 
This information is user's ID and it is shared through BLE using beacon mode (see Figure \ref{ar.fig.beacon}).
Every person has a list of identifiers (IDs) corresponding to nearby people, this list is published in a smart contract so that the participants can be verified.
% MOD <--

The use of BLE allows generating lists containing peers \textit{IDs} that may used when P2P communications is required. When a member of the emergency service receives a new \textit{ID}, he/she checks if it is included in the smart contract list. It the \textit{ID} is not in it, it is not yet included in the contract. Otherwise, it is accepted and stored. 

%% MOD --> 5

In order to make this section more understandable Table \ref{ar.tab.notations} has been added to define some notations. 

\begin{table}[!ht]
\begin{center}
\caption{Notations used}
      \begin{tabular}{cc}
        \hline 
         \textbf{Notation}  &  \textbf{Meaning}  \\ \hline 
        msk  & master secret key \\ 
        mpk  & master public key \\
        $H_i$ & cryptographic hash function (digest) \\
        $H_i(x||y)$ & digest of the concatenation of bit-strings $x$ and $y$ \\
        $x\stackrel{r}{\leftarrow}X$ & $x$ is selected uniformly at random from set $X$ \\ 
        $x \oplus y$ &  bitwise XOR of bitstrings $x$, $y$ of equal length \\
        $\hat{e}(x,y)$ &  bilinear map paring of $x$ and $y$ \\  
        $ x \cdot P $ &  multiplication of the point $P$ times $x$ \\  
        $ x \stackrel{r}{\leftarrow} S$ &  stands for an element $x$ randomly selected from a set $S$ \\
        $x \leftarrow y$ & assignment of the value $y$ to $x$  \\
        $a||b$ & is used for concatenation of $a$ and $b$  \\

    \label{ar.tab.notations}
      \end{tabular}
  \end{center}
\end{table}

% MOD <--

\subsection{Initialization}

    At the end of this stage, all the staff participating in the event must be registered, regardless of the organization to which they belong.

Some elements and basic notation necessary for a detailed description of the system are described below.
    
    \[  H_1: \{0, 1\}^* \rightarrow G^*, H_2 : \{0, 1\}^* \rightarrow \mathbb{Z}^*_q, H_3 : \mathbb{Z}^*_q \rightarrow \{0, 1\}^n, 	\]
    \[  H_4 : \{0, 1\}^n \rightarrow \{0, 1\}^{|m|}, H_5 : G \times G \times \{0, 1\}^n \times {Z}^*_q \times {Z}^*_q \times ... \times {Z}^*_q \rightarrow {Z}^*_q \]
    where $n$ is the length of the message $m$ and $q$ is a prime number. 

    %The steps needed for the signcryption scheme are the following: 
    This initialization phase is carried out in each organization for each person in the system. 
	
    \begin{itemize}

        \item Setup: In this first step, the server initializes the parameters in order to generate its own keys: master public key ($mpk$) and  master secret key ($msk$). This server plays the role of  Private Key Generator (PKG).
        To achieve it, some private data are necessary: $k \in \mathbb{Z}$ to generate a prime $q$ based on it, two groups $\mathbb{G}$ and $\mathbb{G_T}$ of order $q$ and a bilinear pairing map $\hat{e}: \mathbb{G} \times \mathbb{G} \rightarrow \mathbb{G_T}$ are selected. Next, $P \in \mathbb{G}$ is randomly chosen and five hash functions are also defined.   
        
        Finally, server keys are generated: $ msk \stackrel{r}{\leftarrow} \mathbb{Z}^*_q $ and $ mpk \leftarrow msk \cdot P $.
    
        \item After Extract:
    
        In this step, staff identification is carried out.
        Public key $Q_{ID} \in G$ is generated through a hash function applied on the corresponding ID, $ Q_{ID} \leftarrow H_1(ID) $.
    
        Private key $S_{ID}$,  used for communications with the server $S_{ID} \in G$, is calculated taking into account the $msk$, $ S_{ID} \leftarrow msk \cdot Q_{ID} $.
        In the proposal, key exchange between server and staff is done using the stream cipher SNOW3G under the session key obtained through an Elliptic Curve Diffie-Hellman (ECDH) scheme.
    
    \end{itemize}

\subsection{Event generation}

    This phase is carried out by the organization of the user that  generated the alert. The information is stored in the blockchain. 
    %\textit{EVENT EXTRACT:}
    Each one of the generated events  has a unique identifier, $ ID_e \stackrel{r}{\leftarrow} \mathbb{Z}^*_q $ and some location coordinates, $lat$ and $lon$.
    In this stage, the public key for this event $Q_{IDe} \in G$ is generated as: 
    
    \[Q_{IDe} \leftarrow H_1(ID || ID_e || lat || lon) \]
    
    Then, the secret key for this event corresponds with $ S_{IDe} \leftarrow msk \cdot Q_{IDe} $.
    %% --> MOD 5
    To share that private information a secure channel is created, specifically in this system the Elliptic Curve Diffie-Hellman (ECDH) was implemented.
    This protocol is a variation of the original Diffie-Hellman protocol, which uses the properties of the elliptic curves defined over finite fields.
    
    In this way, the operation of this protocol is based on the fact that two users agree beforehand on the use of a prime number $p$, an elliptic curve $E$ defined over $Z_p$ and a point $P \in E$.
    Then, the users A and B, in this case a doctor (A) and the server (B), choose as secret keys two random numbers belonging to $Z_p$. 
    The doctor select $a \in Z_p$ and the server $b \in Z_p$, this are the secret keys $Sk_a$ and $SK_b$ of them.
    Later, both of them obtain their public keys multiplying their secret keys by previously agreed point $P$, obtaining the shared key $SK$. See Table \ref{ar.tab.ecdh} for the specification of the shared key generation.
    
    \begin{table}[!h]
        \caption{Data information related to steps}
        \centering
        % \tablesize{} %% You can specify the fontsize here, e.g.  \tablesize{\footnotesize}. If commented out \small will be used.
        \begin{tabular}{ccc}
        \toprule
        \textbf{Steps}	& \textbf{Medical staff}	& \textbf{Entity server}\\
        \midrule
         0: Initialization	& $p, E, P, Z_p$	& $p, E, P, Z_p$	\\
         1: Secret key generation & $Sk_a \xleftarrow{r} a \in Z_p$	& $Sk_b \xleftarrow{r} b \in Z_p$	\\
         2: Public key generation & $Pk_a \xleftarrow{} a \cdot P $	& $Pk_b \xleftarrow{} b \cdot P$\\
         3: Information exchange & $Pk_b$& $Pk_a$\\
         4: Shared key generation & $K_a = a \cdot Pk_b$ & $K_b = b \cdot Pk_a$\\
         5: Confirmation & $SK == K_a $ & $ SK == K_b$\\
         \bottomrule
         \label{ar.tab.ecdh}
        \end{tabular}
    \end{table}
    
    In order to compute the shared key, the next step is the exchange between them their public keys by multiplying their private key by the public key of the other user/server, obtaining both the same shared key.
    %% <-- MOD 5
    
\subsection{P2P communication scheme in an event}

    The proposed scheme can be used in direct communications between two users.
    %% MOD --> 5 
    The main use of case of this mode is the communication between two doctors in a direct way. For example, if a doctor has a question about a specific point of the emergency or something related to a victim, they can establish a communication channel with this P2P mode.  
    %% MOD <-- 5
    In such a case, the steps are the following:  
    
    \begin{itemize}
    
    	\item Single Signcryption:
    
        All the messages $m \in\{0,1\}^n$ are encrypted and signed.
        Receiver's public key is generated taking into account his/her identification and the pre-shared data ($ID_e$, $lat$ and $lon$): $ Q_{IDe_b} \leftarrow H_1(ID_b || ID_e || lat || lon) $.
        Then, some operations are developed giving as result $\sigma$ (a t-uple of three components: $c$, $T$, $U$). $T$ is generated as $ x \stackrel{r}{\leftarrow}\mathbb{Z}^*_q$ and $ T \leftarrow x \cdot P$.
        Then the signature using sender's private key ($S_{IDe_a}$) is denoted as $U$. It is obtained as follows $ r \leftarrow H_2(T || m)$, $ W \leftarrow x \cdot mpk$ and $ U \leftarrow r \cdot S_{IDe_a} + W $. 
        Finally, the encrypted message is denoted as $c$, and it is generated as $ y \leftarrow \hat{e}(W, Q_{IDe_b}) $, $ k \leftarrow H_3(y)$, $ c \leftarrow k \oplus m$. 
    
    	\item Single Unsigncryption:
    	
        First of all, sender's  public key is generated taking into account $IDe_a$ and the pre-shared information as:
        \[ Q_{ID_a} \leftarrow H_1(IDe_a || ID_e || lat || lon_e) \]
    
        Then $\sigma$ is parsed as $(c, T, U)$. 
        If everything is right, the message $m \in \{0,1\}^n$ is returned. Otherwise, if any problem in the signature or in the encryption of $m$ is detected, $\bot$ is returned.
        The verification consists of :   
        \[ \hat{e}(U, P) == \hat{e}(Q_{IDe_a}, mpk)^r \cdot \hat{e}(T, mpk) \]
        Thus, the user calculates:
        \[ y \leftarrow \hat{e}(S_{IDe_b}, T), k \leftarrow y, m \leftarrow k \oplus c,  r \leftarrow H_2(T || m) \]
    
    \end{itemize}

\subsection{Broadcast communication scheme in an event}

    The proposed scheme can be used when someone wants to share a message.
    %% MOD --> 5 
    In this case, the main use of case of this mode is the communication with everybody around a point. For example, if a doctor want to share some planning order or some alert from their location, he/she can generate a broadcast message by the use of this mode.  
    %% MOD <-- 5
    In such cases, the steps are the following: 

    \begin{itemize}
    
    	\item Broadcast Signcryption:
        In the broadcast mode, there are $n$ receivers, so the sender is identified by $IDe_a$ and the receivers by:
        \[IDe_1, IDe_2, ... , IDe_n\] 
    
        All the broadcast messages $m \in\{0,1\}^n$ are encrypted and signed.
      Sender's public key is generated as: $ Q_{IDe_{a}} \leftarrow H_1(ID_a || ID_e || lat || lon) $.
        Then, some operations are developed, giving as result $ \sigma$ (a t-uple of components: $ c, T, U, V, W, X, a_0, ... a_n-1$). Then, the sender selects some random numbers $ r \stackrel{r}{\leftarrow}\mathbb{Z}^{*}_q$, $ r' \stackrel{r}{\leftarrow}\mathbb{Z}^{*}_q$, $ s \stackrel{r}{\leftarrow}\mathbb{Z}^{*}_q$ and $ p \stackrel{r}{ \leftarrow } \mathbb{Z}^{*}_{q} $ and then, it operates:
        \[T \leftarrow r \cdot Q_{IDe_{a}}, U \leftarrow r \cdot P, X \leftarrow r' \cdot T, J \leftarrow r' \cdot mpk \]
     
        Receivers' public keys are generated taking into account all the identifications $ID_1, ID_2, ... , ID_n$, as follows:
    
        \[f(x) = \prod_{i=0}^{n} (x - v_i) + p (mod q) = a_0 + a_1x + ... + a_{n-1}x^{n-1} + x^n \]
    
        With $ Qe_i \leftarrow H_1(ID_i || ID_e || lat || lon) $, $y_i \leftarrow \hat{e}(Q_i, J)$ and $ v_i \leftarrow H_2(y_i)$. 
        Then it calculates $V \leftarrow s \cdot H(p)$, the key $k$ as $ k \leftarrow H(s) $ and the encrypted message $c$ as $c \leftarrow k \oplus m $.  
        Finally an authenticator $h$ is generated as $ h \leftarrow H_5(c, X, U, V, a_0, a_1, ..., a_{n-1})$ and 
        $ W \leftarrow ( r' + h ) r \cdot S_{ID_a}$. 
    %$ \cdot Q_{IDe_{a}}$ and $U$ with $ \alpha \stackrel{r}{\leftarrow}\mathbb{Z}^*_q$ as $ U \leftarrow \alpha \cdot Y$. 
    %$ c \leftarrow \beta \oplus H_3(p)$ and $W$ with $ \beta \stackrel{r}{\leftarrow}\mathbb{Z}^*_q$, $ K \leftarrow H_4(\beta) $, $ Z \leftarrow Snow3G_K(m) $,
    
        \item Multiple Receiver Unsigncryption:
    
        In this step, two verifications are carried out, but first of all, 
        $\sigma$ is parsed as $ c, T, U, V, W, X, a_0, ... a_n-1 $ and $  h \leftarrow H_5(c, X, U, V, a_0, a_1, ..., a_{n-1})$. 
        The first verification is the public verification to check that the ciphertext is valid:
        
        \[ \hat{e}(W, P) == \hat{e}(X + hT, mpk) \]
        Otherwise, the ciphertext has been damaged or  is invalid, and $\bot$ is returned. 
        The second verification is:
    
        \[\hat{e}(W, Qe_i) == \hat{e}(X + hT, S_{IDe_i}) \]
        
        This is to check if $ID_i$ is one of the receivers chosen by the
        sender and if the ciphertext is valid. Otherwise, the receiver  quits the decryption process and $\bot$ is returned. 
        To generate the message, some operations are generated: $ y_i \leftarrow \hat{e}(S_{IDe_b}, U)$, $ v_i \leftarrow H_2(y_i)$,
        $ p \leftarrow f(v_i)$, $ s \leftarrow V \oplus H_3(p) $, $ k \leftarrow H_4(s) $ and $ m \leftarrow k \oplus c$.
    
    \end{itemize}
%AQUÍ

%**************************************************************************************
\section{Implementation and analysis}\label{ar.sec.anali}

A first implementation of the proposal has been developed to obtain a prototype as a proof-of-concept. The smart contract has been implemented using Solidity (Ethereum coding language) on a permissioned blockchain where only authorized entities are assigned writing permission.

\begin{figure}[!ht]
    \centering
    \includegraphics[width=1\linewidth]{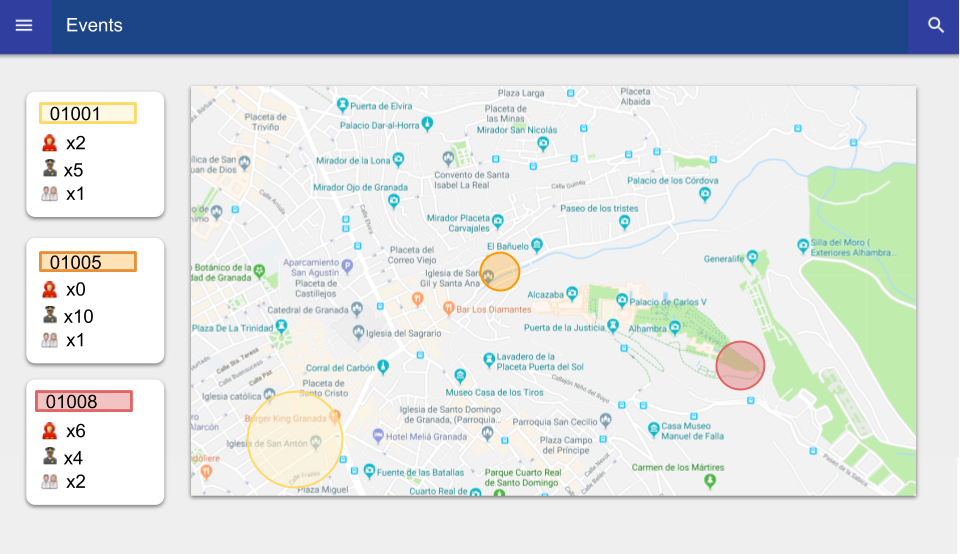}
    \caption{Web application.} \label{ar.fig.web}
\end{figure}

This approach allows emulating a permissioned blockchain without the need to spend real money.
In this prototype, the user interface is managed by a decentralized web application (DApp), which allows user interaction with the smart contract on the blockchain (see Figure \ref{ar.fig.web}).

%% MOD --> 4 

Moreover, an exhaustive analysis about the performance of the BLE communications has been done. 
In this case, using the Opportunistic Network Environment simulator (ONE) \cite{keranen-theone} a total of 300 people have been randomly deployed over an area of $2km^2$. 
This tool allows selecting different communication technologies, interfaces, node behaviour, node speed, node deployment, simulation time, etc. 
Using the different parameters available on the tool the behaviour of BLE technology in this scenario has been simulated.

\begin{figure}[!ht]
    \centering
    \includegraphics[width=1\linewidth]{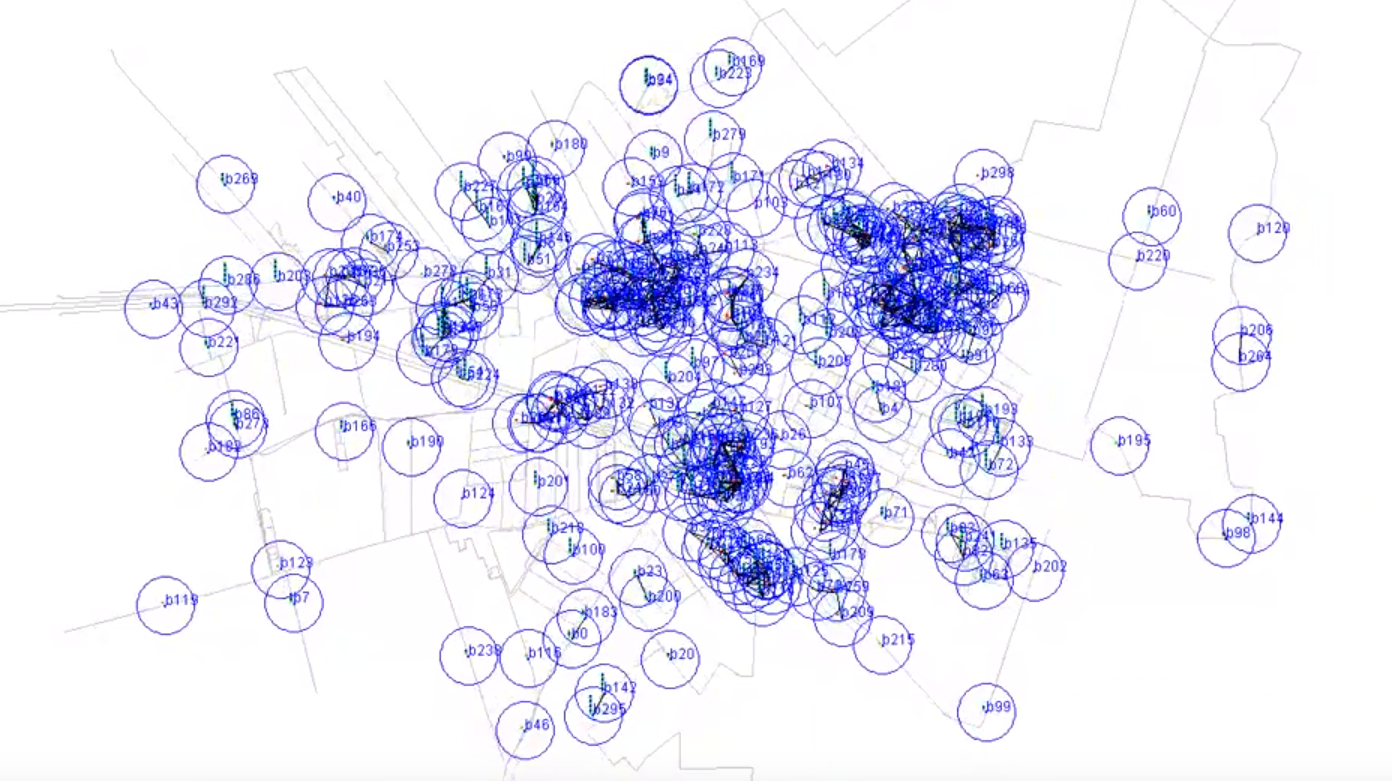}
    \caption{BLE simulated scenario.} \label{ar.fig.btSimu}
\end{figure}

In Figure \ref{ar.fig.btSimu}, the range and communication links of the BLE technology in the simulated area can be observed. 
Moreover, to obtain enough and precise data about the BLE technology in this kind of situations, the simulation has been repeated 10 times. An important measurement that have to be taken into account in this kind of scenarios is the average of communications reached. 
This measure allows us to know how many nodes have received  communications by part of the rest of the nodes of the system, fact that is important because if this measurement is extrapolated we can know the the isolated nodes of the network.

The other measurement taken is the average of communications received by a node. 
This is an important measurement because it allows to know haw many communications receive in average every node of the system during the simulation time. 

\begin{table}[!ht]
	\begin{center}
	\begin{tabular}{| c | c |}
	\hline
	Communications reached & 2646.2 \\
    \hline
  	Isolated nodes & 17.4 \\
    \hline
	Communications received by node & 8.82 \\
	\hline
	
	\end{tabular}
	\end{center}
	\caption{BLE communication results in the simulated scenario.}
	\label{tab:BTSim}  
\end{table}

The obtained results can be observed in Table \ref{tab:BTSim}, where the an amount of 2646.2 communications of the system have been reached in average. Moreover, during the simulated time 17.4 nodes were isolated supposing it only the 5.6\% of the communications. 
Finally, an amount of 8.82 communications have been received by each node of the system in average during the simulated scenario, supposing it that every node of the system will communicate or will be communicated every 7 minutes approximately.

%% MOD <-- 4 

Security is one of the priorities in order to protect the system against attacks such as Denial of Services (DoS), Man in the Middle (MitM) and impersonation.
DoS attacks are limited because only requests associated with a legitimate number of members of the Emergency Services take effect.
On the other hand, the typical MitM attack, which conveys a successful authentication to the server with a legitimate identifier is very improbable, since once the legitimate user private key is assigned to the server, further requests of this kind will not be taken care of.
Impersonation will be easily detectable because the number of members who can make requests to the server is limited to those who are working at the time of the request.

% MOD --> 4
An analysis of efficiency related to the technologies coverage, their range and their transmission efficiency was developed.
A beta prototype has been also implemented with Wi-Fi Aware but in the preview mode of the technology.

% MOD <-- 4

% MOD --> 1
The elements of the blockchain are replicable on different computers, since what underlies  is a distributed database. This justifies that it is impossible to falsify its elements, since all the data are generally found in several servers, and its synchronization occurs almost simultaneously. In addition, if a falsification is achieved with one of the registers, it should be easy to detect it through the codes that link the blocks and prevent it from spreading.
    
Before including a transaction into a block it is always verified. This verification is carried out by miners, which are computers dedicated not only to keep a copy of the data but also to validate nodes' agreement. 
% MOD <-- 

Thus, since the validators of this kind of blockchain are known, there is no risk of successful 51\% attack.

% MOD --> 1 
    Smart contracts are executed and enforced automatically and autonomously without the intervention of third parties.
    Whats more, a smart contract can be created and executed by individuals and/or legal entities, but also by machines or other programs that work autonomously.

    Due to its nature, a smart contract is valid without the need of authorities because it is a code. Blockchain technology makes possible to share this code with all network nodes, guaranteeing that it cannot be modified. Hence, the main features of smart contracts are decentralization, persistence and transparency.
 
    The so-called oracles are  elements that allow the smart contract to interact with the real world. They are autonomous tools that allow updating the internal states of a smart contract through external information, usually obtained with APIs.
    In the proposal here described, oracles are smart devices used by the emergency services.

    The oracles also work autonomously. 
    However, it must be taken into account that the source used by the oracle is a third party that must be trusted, and that could be corrupted by its owner, crack, or could simply fail your server, something that has negative implications.
% MOD --<  

\begin{figure}[!ht]
    \centering
    \includegraphics[width=0.8\linewidth]{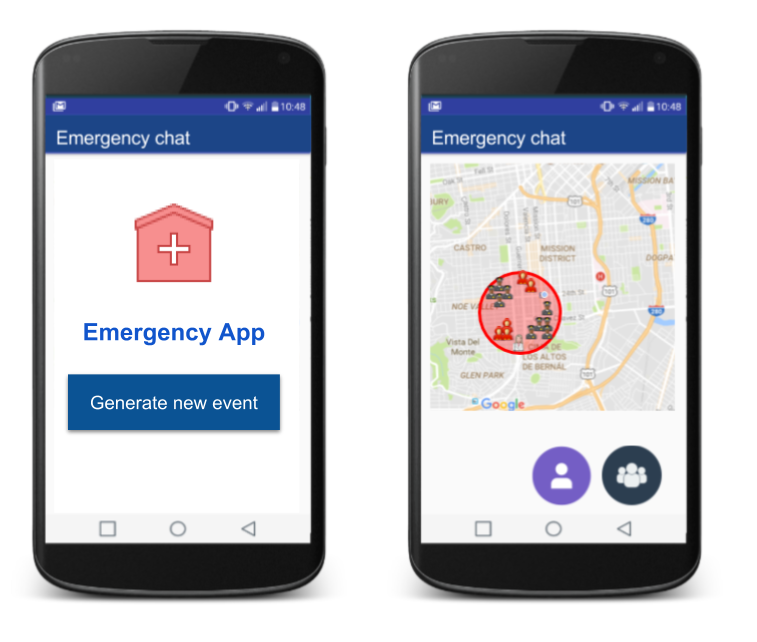}
    \caption{Developed prototype.} \label{ar.fig.app}
\end{figure}

% MOD -->
An Android mobile application was developed to improve communication between emergency services in extreme situations (see Figure \ref{ar.fig.app}). 
This implementation has been tested tested through the generation of some random events located on a map and assigning ad-hoc users to generated events. Smart contracts were developed with Solidity \cite{solidity}, and deployed using Truffle Suite \cite{truffle}. 
On the server side, we use nodejs \cite{node} with WEB3JS \cite{web3js} to connect with the smart contracts trough Ganache \cite{ganache} and Drizzle \cite{drizzle} on the front-end. 
% MOD <--

The  prototype requires communication with the PKG only at the initialization stage where the key generation is performed.
Afterwards, users can share messages with their own keys and with the pre-shared information related to the event.
An alternative scheme with direct communications without requiring central server is deployed. Note that this approach can be used when communications are saturated.

\section{Conclusions and future work}

In this paper, a decentralized low-cost model has been proposed, which supports the monitoring of risk levels in big events, aiming at improving resource allocation and staff coordination in case of emergency.

The proposed system is based on blockchain and on the establishment of an alternative communication channel for sharing information while the emergency situation. An important feature of this system is the no need for cloud services.

The system generates automatically the pre-shared data related to the event to which the staff is assigned.

The proposal includes a web application used to manage all  emergency services and incidents that are generated and publicly notified through its inclusion  in a  smart contract and in the blockchain. Hence, once the information regarding an event is included in this structure, anybody may consult it and its validation remains guaranteed.  

A mobile application with an ubiquitous Wi-Fi Direct chat has been implemented where communication security is based on the use of IDentity-Based Cryptography. BLE in beacon mode is used to support identity exchange among participants.
Emergency services are able to know, through the mobile application, where the event is located and where they must be deployed, as well as their peers location.

The main objective for which an alternative communication channel is proposed is that emergency personnel can share information, either in P2P or broadcast mode, when an incident at a big event saturates the network.
ID-Based Signcryption has been included to guarantee integrity, confidentiality, authenticity and non-repudiation in communications.
A beta prototype has been implemented with Wi-Fi Aware, which is available only on Android 8 and in the preview mode of this technology.
The possible incorporation of LTE-Direct depends on the Native Development Kit because at this moment this code is private. 
As part of work in progress, a real test with emergency staff is planned so that the smart contract can change to improve the model according to the results.
As future work, more functionalities will be added to the server, such as statistics, private chats based on roles, etc.
In the short term,  the system modeling and the smart contract definition are expected to be improved  on the basis of the results of real field tests to be developed involving emergency personnel.
The possibility of adding new functionalities to the system, such as obtaining and analysing relevant statistics for emergency management and the definition of private chats  based on  roles are issues under study.

\section*{Acknowledgments}

This research was funded by the Centre for the Development of Industrial Technology (CDTI) under project C2017/3-9 (UNICRINF), by the Spanish Ministry of Science, Innovation and Universities (MCIU), the State Research Agency (AEI) and the European Regional Development Fund (ERDF) under project RTI2018-097263-B-I00 (ACTIS) and by the Government of the Canary Islands through TESIS2015010102 and TESIS-2015010106 grants.

%\subsection*{Author contributions}
%This is an author contribution text. This is an author contribution text. This is an author contribution text. This is an author contribution text. This is an author contribution text. 

%\subsection*{Financial disclosure}
%None reported.

\subsection*{Conflict of interest}

    The authors declare no potential conflict of interests.

%\section*{Supporting information}
%The following supporting information is available as part of the online article:

%\noindent
%\textbf{Figure S1.}
%{500{\uns}hPa geopotential anomalies for GC2C calculated against the ERA Interim reanalysis. The period is 1989--2008.}

%\noindent
%\textbf{Figure S2.}
%{The SST anomalies for GC2C calculated against the observations (OIsst).}

%\nocite{*}% Show all bib entries - both cited and uncited; comment this line to view only cited bib entries;
\bibliographystyle{NJDnatbib}
\bibliography{arBiblio}

\begin{thebibliography}{10}
\providecommand \doibase [0]{http://dx.doi.org/}%

\bibitem{gupta2016inside}
Panwar G, Misra S. Inside Bluetooth Low Energy (Gupta, {N.)} [Book Review]. In:
  . 24. ; 2017\string: 2--3.

\bibitem{shen2016secure}
Shen W, Yin B, Cao X, Cai LX, Cheng Y. Secure device-to-device communications
  over WiFi direct. In: . 30. ; 2016\string: 4--9.

\bibitem{ar.Boneh2001}
Boneh D, Franklin MK. Identity-Based Encryption from the Weil Pairing. In: .
  32. ; 2003\string: 586--615.

\bibitem{malone2002identity}
Malone{-}Lee J. Identity-Based Signcryption. In: . 2002. ; 2002\string: 98.

\bibitem{Sharmila2008Efficient}
Selvi SSD, Vivek SS, Srinivasan R, Rangan CP. An Efficient Identity-Based
  Signcryption Scheme for Multiple Receivers. In: ; 2009\string: 71--88.

\bibitem{das2019radio}
Das D, Huang PK, Elad O, Qi EH, Park M. Radio resource allocation in wi-fi
  aware neighborhood area network data links. 2019.
\newblock US Patent App. 16/141,313.

\bibitem{android2011android}
documentation dA. {Wi-Fi Aware on Android}. 2019.

\bibitem{dOroCGIMP19}
d'Oro EC, Colombo S, Gribaudo M, Iacono M, Manca D, Piazzolla P. Modeling and
  evaluating a complex edge computing based systems: An emergency management
  support system case study. {\it Internet of Things} 2019\string; 6.
\newblock \href {\doibase 10.1016/j.iot.2019.100054} {doi:
  10.1016/j.iot.2019.100054}

\bibitem{motta2010wireless}
Motta R, Pasquale J. Wireless {P2P}: Problem or opportunity?. In: ;
  2010\string: 32--37.

\bibitem{conti2013experimenting}
Conti M, Delmastro F, Minutiello G, Paris R. Experimenting opportunistic
  networks with WiFi Direct. In: ; 2013\string: 1--6.

\bibitem{camps2013device}
Camps-Mur D, Garcia-Saavedra A, Serrano P. Device-to-device communications with
  {Wi-Fi Direct}: overview and experimentation. In: . 20. IEEE; 2013\string:
  96--104.

\bibitem{santos2014alternative}
Santos{-}Gonz{\'{a}}lez I, Rivero{-}Garc{\'{\i}}a A, Caballero{-}Gil P,
  Hern{\'{a}}ndez{-}Goya C. Alternative Communication System for Emergency
  Situations. In: ; 2014\string: 397--402.

\bibitem{rivero2018secure}
Rivero-Garc{\'i}a A, Santos-Gonz{\'a}lez I, Goya CH, Caballero-Gil P. Secure
  Communication System for Emergency Services in Network Congestion Scenarios.
  In: ; 2018\string: 63.

\bibitem{nakamoto2008bitcoin}
Nakamoto S. Bitcoin: A peer-to-peer electronic cash system.. In: Working Paper;
  2009.

\bibitem{ekblaw2016case}
Ekblaw A, Azaria A, Halamka JD, Lippman A. A Case Study for Blockchain in
  Healthcare: {MedRec} prototype for electronic health records and medical
  research data. In: . 13. ; 2016\string: 13.

\bibitem{dubovitskaya2017secure}
Dubovitskaya A, Xu Z, Ryu S, Schumacher M, Wang F. Secure and Trustable
  Electronic Medical Records Sharing using Blockchain. In: ; 2017.

\bibitem{Radanovic2018}
Radanovi{\'{c}} I, Liki{\'{c}} R. Opportunities for Use of Blockchain
  Technology in Medicine. In: ; 2018.

\bibitem{MinoliO18}
Minoli D, Occhiogrosso B. Blockchain mechanisms for IoT security. {\it Internet
  of Things} 2018\string; 1-2\string: 1--13.
\newblock \href {\doibase 10.1016/j.iot.2018.05.002} {doi:
  10.1016/j.iot.2018.05.002}

\bibitem{cha2018blockchain}
Cha S, Chen J, Su C, Yeh K. A Blockchain Connected Gateway for BLE-Based
  Devices in the Internet of Things. In: . 6. ; 2018\string: 24639--24649.

\bibitem{guo2018secure}
Guo R, Shi H, Zhao Q, Zheng D. Secure Attribute-Based Signature Scheme With
  Multiple Authorities for Blockchain in Electronic Health Records Systems. In:
  . 6. ; 2018\string: 11676--11686.

\bibitem{yue2016healthcare}
Yue X, Wang H, Jin D, Li M, Jiang W. Healthcare Data Gateways: Found Healthcare
  Intelligence on Blockchain with Novel Privacy Risk Control. In: . 40. ;
  2016\string: 218:1--218:8.

\bibitem{griggs2018healthcare}
Griggs KN, Ossipova O, Kohlios CP, Baccarini AN, Howson EA, Hayajneh T.
  Healthcare Blockchain System Using Smart Contracts for Secure Automated
  Remote Patient Monitoring. In: . 42. ; 2018\string: 130:1--130:7.

\bibitem{keranen-theone}
Ker\"{a}nen A, Ott J, K\"{a}rkk\"{a}inen T. {The ONE Simulator for DTN Protocol
  Evaluation}. In: ICST; 2009; New York, NY, USA.

\bibitem{solidity}
Ethereum . Solidity. \url{https://solidity-es.readthedocs.io/es/latest/#};
  2019.
\newblock Accessed: Thu, 24 Sep 2019.

\bibitem{truffle}
Suite TBGT. Truffle Suite. \url{https://www.trufflesuite.com/};  2019.
\newblock Accessed: Thu, 24 Sep 2019.

\bibitem{node}
Foundation N. Node.js. \url{https://nodejs.org};  2019.
\newblock Accessed: Thu, 24 Sep 2019.

\bibitem{web3js}
Ethereum . Web3.js. \url{https://web3js.readthedocs.io/en/1.0/};  2019.
\newblock Accessed: Thu, 24 Sep 2019.

\bibitem{ganache}
Ganache TBG. Ganache. \url{https://github.com/trufflesuite/ganache};  2019.
\newblock Accessed: Thu, 24 Sep 2019.

\bibitem{drizzle}
Drizzle TBG. Drizzle. \url{https://github.com/trufflesuite/drizzle};  2019.
\newblock Accessed: Thu, 24 Sep 2019.

\end{thebibliography}

\end{document}